\documentclass{aa}
\usepackage{graphicx}

\begin{document}
\thesaurus{08               
                 (09.19.2,  
		 13.25.3,  
		 13.18.2,  
		 13.07.3)  
}
\title{Thermal X-ray composites as an effect of projection} 
\author{O.~Petruk
	}
\institute{Institute for Applied Problems in Mechanics and Mathematics, 
	   3-b Naukova St., Lviv 79053, Ukraine \\
	   {petruk@astro.franko.lviv.ua}}
\date{Received ... ; Accepted ...}


\abstract{
A new possibility to explain the nature of thermal X-ray composites 
(TXCs), i.e. a class of supernova remnants (SNRs) with a thermal X-ray 
centrally-filled morphology within a radio shell, as a projection effect of 
the 2- or 3-dimensional 
shell-like SNR evolv\-ed in a nonuniform medium 
with scale-height $\le10\ {\rm pc}$ is proposed. 
Both X-ray and radio morphologies, as well as the basic 
theoretical features of this kind of SNR and the surrounding medium, 
are considered. 
Theoretical properties of a shell-like SNR evolved at the edge of a 
molecular cloud correspond to the observed properties of TXCs 
if the gradient of the ambient density does not lie in the projection plane 
and the magnetic field is nearly aligned with the line of sight. 
So, at least a part of objects from the class may be interpreted 
within the framework of the considered effect. 
The proposed model suggests that SNRs with barrel-like radio 
and centrally-brightened thermal X-ray morphologies should exist. 
The model allows us to consider TXCs as prospective sources of proton 
origin $\gamma$-rays.  
}
\maketitle

\section {Introduction}
	\label{intro}

Observations show that supernova remnants (SNRs) have 
anisotropic distributions of surface brightness 
(Seward \cite{Seward-catalog}; Whiteoak \& Green \cite{MOST-catalog}). 
There are four morphological classes of SNRs: shell-like, Crab-like 
(plerionic), composite and thermal X-ray composites (TXCs; 
or mi\-x\-ed-mor\-pho\-lo\-gy, or cent\-ral\-ly-in\-flu\-en\-ced) 
(Jones et al. \cite{Workshop-97}; Rho \& Petre \cite{Rho-Petre-98}, 
hereafter RP98). 
In the past few years interest in TXCs has risen 
(e.g., Jones et al. \cite{Workshop-97}; 
Cox et al. \cite{W44}; Shelton et al. \cite{W44-II}; 
Sun, Wang \& Chen \cite{SunWangChen-99}). 
TXCs are SNRs with centrally concentrated thermal X-ray and limb 
brightened radio morphologies. 
Remnants \object{W44}, \object{W28}, \object{3C~400.2}, \object{Kes~27}, 
\object{3C~391}, \object{CTB~1}, \object{MSH~11-61A} 
and others represent a mixed-morphology class (RP98). 
Since the publication of RP98, a number of new SNRs have been discovered, 
others have been observed more precisely or even for the first time 
in X-rays. 
Thus, new candidates to the TXC class have appeared in the past two years. 

Two physical models have been presented so far to explain TXC 
(see RP98 for review). One of them 
is an enhanced interior X-ray emission from the 
evaporated material of numerous swept-up clouds which 
increases density in the central region of SNR. 
This model is frequently used; sometimes its application is intrinsically 
inconsistent, e.g., as in MSH~11-61A where evaporation timescales exceed the 
age of SNR 50-100 times (Jones et al. \cite{Workshop-97}). 
In the second model, shock temperature is small due to 
essential cooling; very soft emission of the shell is absorbed by 
interstellar medium (ISM) and only the interior region remains visible. 
In this model, thermal 
conduction may level temperature profiles and increase the 
central density altering the interior structure (Cox et al. \cite{W44}). 

Other possibilities are also noted in publications. 
They are: a) emission from ejecta, b) differential absorption 
(Long et al. (\cite{Long-et-al-91}) have concluded 
that the centrally-peaked X-ray morphology within a radio shell 
is unlikely the result of absorption alone because 
distributions of X-ray and radio emitting plasmas have to be 
different in this case) or c) explosion in a medium with 
centrally-concentrated density distribution  
$\rho(r)\propto r^{-m}$, $m>0$ (such a medium does not 
give centrally-concentrated morphology (Long et al. \cite{Long-et-al-91}) 
because the Sedov (\cite{Sedov}) solutions give a specific internal profile 
of the flow density for all 
$m\leq 2$ ($\gamma=5/3$), other $m$ develop a cavity around the centre%
).

The mentioned models are used to obtain a centrally-filled morphology 
within the 
framework of one-di\-men\-si\-o\-nal (1-D) hydrodynamic approaches. 
When we proceed to 2-D or 3-D models, we note that a simple projection 
effect 
may cause the shell-like SNR to fall into another morphology class, 
namely, centrally-influenced (Hnatyk \& Petruk \cite{Hn-Pet-99}). 
The main feature of such SNR is the thermal X-rays 
emitted from swept-up gas and 
peaked in the internal part of the projection. 
Therefore, getting beyond one dimension, we obtain a new possibility 
to explain the nature of TXC. 
Such a possibility is considered in this paper. 

Hnatyk \& Petruk (\cite{Hn-Pet-98}) 
have reproduced the X-ray morphology of IC~443 
with the proposed projection model of TXC. 
Therefore, we restrict ourselves to theoretical consideration 
of the phenomenon. 

SNRs are modelled with an approximate analytic me\-thod for  
hydrodynamic description of the adiabatic phase of an asymmetrical point 
explosion in an arbitrary 
large-scale nonuniform medium (Hnatyk \& Petruk \cite{Hn-Pet-99}). 
Equilibrium thermal X-ray emission is calculated with the use 
of the Raymond \& Smith (\cite{Raym-Smith77}) model. 
The model for radio emission of nonspherical SNR is 
based on Reynolds \& Chevalier (\cite{Rey-Chev-81}) and 
Reynolds (\cite{Reyn-98}) and is described in Sect.~\ref{emis}. 

\section {Observed properties of thermal X-ray composites}
	\label{Sect-prop}

RP98 prove that mi\-xed-mor\-pho\-lo\-gy 
SNRs create a separate morphology class since their properties distinguish 
these remnants from others. Having analysed X-ray data 
on a number of such SNRs, the authors found their two prominent 
morphological distinctions: 
a) the X-ray emission is thermal, the 
distribution of X-ray surface brightness 
is centrally-peaked or amorphous and fills the area within the radio shell 
and may reveal weak evidence of an X-ray shell, 
b) the emission arises primarily from the swept-up ISM material, not from 
the ejecta. 

RP98 emphasize that, besides similar morphology, the sample 
of SNRs also has similar physical properties. Na\-me\-ly, 
a) the same or higher central density comparing with the edge, 
b) complex interior optical nebulosity, as in \object{W28} and probably in 
\object{3C~400.2} (Long et al. \cite{Long-et-al-91}); 
c) higher emission measure $\int n_e^2dl$ ($n_e$ is the 
electron number density, $l$ is the length inside SNR) 
in the central region, as e.g. in \object{3C~391} 
(Rho \& Petre \cite{Rho-Petre-96}); 
d) X-ray surface brightness in the central region 
($r<0.2R_{\rm p},\ R_{\rm p}$ is 
the average radius of the projection), in general, exceeds the brightness 
near the edge ($r>0.6R_{\rm p}$) 2-5 times,
e) temperature profiles are close to uniform. 
As to the latter property, it should be noted that the 
temperature may decrease towards the centre, as in \object{3C~391} 
(Rho \& Petre \cite{Rho-Petre-96}); 
no strong evidence of increasing the temperature towards the centre 
has been found for all TXCs, 
but Cox et al. (\cite{W44}) note that spectral hardness in \object{W44} 
is greater in the centre, so the temperature might be higher 
in this region. 
A possible variation of temperature may be within 
factor 2, as in the case of \object{W44} 
(Rho, Petre \& Schlegel \cite{Rho-et-al-94}; Cox et al. \cite{W44}) or in 
\object{W28} (Long et al. \cite{Long-et-al-91}). 

7 objects from the list of 11 TXCs 
reveal observational evidence 
of an interaction with molecular clouds (RP98). 
Thus, ambient media in the regions of their localization are nonuniform 
and cause nonspherisity of SNRs. 
Observational evidence of cloud localization just on the line of sight 
for some of these SNRs also exists (e.g., Rho et al. \cite{Rho-et-al-94}). 



\section {Theoretical properties of "projected composites"}
	\label{theor-prop}

The Sedov (\cite{Sedov}) model does not give a centrally-concentrated 
morphology due to geometrical properties of self-similar solutions%
. 
The solutions are 1-D and give a specific internal profile 
of the flow gas density: 
most of the mass is concentrated near the shock 
front
. 
These factors and cumulation of the emission along the line of sight cause 
a shell-like morphology. 
If we consider a more complicated nonuniform ISM, 
we get beyond one dimension and need to consider additional parameters 
responsible for nonuniformity of the medium and orientation of 
a 3-D object. 

Projection effects 
may essentially change the morphology of SNRs 
(Hnatyk \& Petruk \cite{Hn-Pet-99}). 
Densities over the surface of a nonspherical SNR may essentially 
differ in various regions. 
If the ambient density distribution 
provides a high density in one of the 
regions across the shell of SNR and is high enough to 
exceed the internal column density near the edge of the projection, 
we will see a centrally-filled projection of a really shell-like SNR. 
Such density distribution may be ensured e.g. by a molecular cloud 
located near the object. 

What is a really shell-like 3-D SNR? We suggest that such a remnant has 
internal density profiles similar to those in the Sedov (\cite{Sedov}) 
solutions. Thus, we separate a shell-like SNR (as an intrinsic property 
of a 3-D object) from its limb 
brightened projection (as a morphological property of the projection). 
Let us call shell-like SNRs with centrally-filled 
projections "projected composites". 

\subsection{Hydrodynamic models}

For simplicity, let us consider the case of a 2-D SNR 
and the characteristics 
of SNR and the surrounding medium which could be possible 
on smoothed boundaries of 
molecular clouds. Thus, SNR evolves in the ambient medium with 
hydrogen number density $n$ distributed according to 
\begin{equation}
n(\tilde{r})=n_o+n_c\exp(-\tilde{r}/h),
\label{eq-1}
\end{equation}
where $n_o$ is the density of the intercloud medium, 
the second term represents the density distribution into the 
boundary region and the cloud, 
$h$ is the scale-hight, $\tilde{r}$ is the distance
. 
Let us take the explosion site to be at point $\tilde{r}_o$ where 
$n(\tilde{r}_o)=2n_o$. 
Other parameters are assumed to be $n_o=0.1\ {\rm cm^{-3}}$, 
$n_c=100\ {\rm cm^{-3}}$%
. 
The energy of the supernova explosion is $E_o=1\cdot 10^{51}\ {\rm erg}$. 
We consider three basic evolutionary cases of SNR models 
which cover practically the whole adiabatic phase (models $a$-$c$, 
Table~\ref{tabl-1}\footnote{The denser part of the 
shell in model $c$ just enters 
the radiative phase since this part has $\lg(T_{\rm s}, {\rm K})=5.6$ 
and the transition temperature for 
the Sedov blast wave is $\lg(T_{\rm s}, {\rm K})=5.8$ 
(Blondin et al. \cite{Blondin-et-al-98}).})
and then we vary 
parameter $h$ in intermediate model $b$ 
(models $d$-$f$, Table~\ref{tabl-1}), 
in order to see how the gradient of the ambient density 
affects X-ray characteristics of objects. 

\begin{table}[t]
\caption[]{
Parameters of SNR models. 
$h$ is the scale-height in the ambient medium density distribution 
(\ref{eq-1}). 
$t$ is the age of the SNR, 
$R$ and $D$ are the radius and velocity of the shock front, 
$T_{\rm s}$ and $n_{\rm s}$ are the temperature and number density 
of the swept-up gas right behind the shock. 
$R_{\rm max}$, $R_{\rm min}$ ($D_{\rm max}$, $D_{\rm min}$) 
are the maximum and minimum shock radii (shock velocities) of 
nonspherical SNR. Analogously, 
$T_{\rm s,\ \!max}$, $T_{\rm s,\ \!min}$ 
($n_{\rm s,\ \!max}$, $n_{\rm s,\ \!min}$) 
are maximum and minimum temperatures (number densities) of the gas flow 
right behind the shock.
$L_{\rm x}^{>0.1\ {\rm keV}}$ 
is the thermal X-ray luminosity (for photon energy $>0.1\ {\rm keV}$) 
and $\alpha$ is the spectral 
index (at photon energy $5\ {\rm keV}$) 
of the thermal X-ray emission from the whole SNR. 
$T_{\rm ef}$ is the effective temperature of a nonspherical SNR 
defined by Hnatyk \& Petruk (\cite{Hn-Pet-99}) as 
$T_{\rm ef}\propto M^{-1}$, where $M$ is the swept-up mass. 
The contrasts in the distribution of X-ray surface brightness 
$\lg (S_{\rm c}/S_{\rm max,\ \!2})$ and spectral index 
$\alpha_{0.95}/\alpha_{\rm c}$ are presented for the case of 
$\delta=90\degr$. Subscript "c" corresponds to the center of the 
projection, 
$\alpha_{0.95}$ is the value of the index at $0.95R_{\rm p}$, 
$R_{\rm p}$ is the radius of the projection. 
	}
\begin{flushleft} 
\begin{tabular}{lcccccc} 
\hline
Parameter&\multicolumn{6}{c}{Model} \\
\cline{2-7}
&$a$&$b$&$c$&$d$&$e$&$f$\\
\hline
$h,\ {\rm pc}$                        & 2.5     & 2.5     & 2.5  & 5    & 10   & 40 \\
$t,\ {\rm 10^3\ yrs}$                 & 1.0     & 6.8     & 17.7 & 6.8  & 6.8  & 6.8 \\
$\lg T_{\rm ef},\ {\rm K}$            & 8       & 7       & 6.5  & 7    & 7    & 7  \\
$M,\ {\rm M_{\sun}}$                  & 9.5     & 94      & 280  & 98   & 95   & 94 \\
$R_{\rm max}/R_{\rm min}$             & 1.4     & 1.8     & 2.1  & 1.4  & 1.2  & 1.1 \\
$D_{\rm max}/D_{\rm min}$             & 1.9     & 2.8     & 3.1  & 1.9  & 1.5  & 1.1 \\
$T_{\rm s,\ \!max}/T_{\rm s,\ \!min}$ & 3.5     & 7.9     & 9.8  & 3.7  & 2.2  & 1.2 \\
$n_{\rm s,\ \!max}/n_{\rm s,\ \!min}$ & 9.5     & 45      & 84   & 11   & 3.9  & 1.4 \\
$\lg L_{\rm x}^{>0.1\ {\rm keV}}$     & 34.1    & 36.7    & 37.3 & 36.4 & 36.2 & 36.1 \\
$\alpha^{5\ {\rm keV}}$               & 0.98    & 3.2     & 3.1  & 3.9  & 4.1  & 4.2 \\
$\lg (S_{\rm c}/S_{\rm max,\ \!2})$   & 0.43    & 2.1     & 2.3  & 0.72 &-0.10 & -0.54 \\
$\alpha_{0.95}/\alpha_{\rm c}$        & 1.6     & 1.8     & 3.6  & 1.3  & 1.2  & 1.3 \\
\hline\end{tabular}
\end{flushleft}
\label{tabl-1}
\end{table}

\subsection{Emission models}
\label{emis}

Gas density $n$ and temperature $T$ distributions inside the 
volume of a nonspherical SNR are obtained with the method of 
Hnatyk \& Petruk (\cite{Hn-Pet-99}).

The equilibrium thermal X-ray emissivities are taken from 
Raymond \& Smith (\cite{Raym-Smith77}). 

We make simple estimations of the 
radio morphology of SNR in a nonuniform medium as described below, 
on the basis of a model for synchrotron emission from SNRs developed by 
Reynolds (\cite{Reyn-98}, hereafter R98) and 
Reynolds \& Chevalier (\cite{Rey-Chev-81}). 

The volume emissivity in the radio band at some frequency $\nu$ is 
\begin{equation}
S_\nu\propto KB_\perp^{(s+1)/2}, 
\end{equation}
where $K$ is the normalization of electron distribution 
$N(E)dE=KE^{-s}dE$, 
$B_\perp$ is the tangential component 
of the magnetic field (perpendicular to the electron velocity or 
normal to the shock). 
Power index $s$ is constant downstream because electrons lose 
energy 
proportionally to their energy 
(Eq.~(\ref{en-loss})) and remain essentially confined to the fluid element 
in which they were produced. 
Power index $s$ is also constant over the surface of a nonspherical 
SNR because the shock is strong. Namely, in the first order 
Fermi acceleration mechanism, $s=(\sigma+2)/(\sigma-1)$, where  
shock compression ratio $\sigma$ does not depend on the ambient density 
distribution in the strong shock limit (Mach number $\gg 1$) and 
$\sigma=4$ for $\gamma=5/3$. 

The ambient field is assumed to be uniform 
(polarization observations 
support the assumption that the magnetic fields 
in molecular clouds may be ordered over large scales, 
e. g. Goodman et al. (\cite{Goodman-et-al-90}), 
Messinger et al. (\cite{Messinger-et-al-97}), 
Matthews \& Wilson (\cite{Matthews-Wilson-2000}) and others). 
Component $B_\parallel$ is not modified by the shock: 
$B_{\parallel,{\rm s}}/B_{\parallel,{\rm s}}^o=1$ 
(indices "$s$" and "$o$" refer to the values at the shock and to the 
surrounding 
medium, respectively). 
Component $B_\perp$ rises everywhere at the shock by the factor of 
$\rho_{\rm s}/\rho^o_{\rm s}=\sigma$. 
No further turbulent amplification of the magnetic field is assumed. 
Components $B_\parallel$ and $B_\perp$ evolve differently behind the 
shock front (R98; Reynolds \& Chevalier \cite{Rey-Chev-81}). 
Since the magnetic field is flux-frozen, 
$B_\perp(r)rdr={\rm const}$, 
the tangential component in each 1-D sector of the remnant is 
\begin{equation}
B_\perp(r)=B_\perp(a)\ {\rho(r)\over\rho(a)}\ {r\over a},
\end{equation}
where $a$ is the Lagrangian coordinate. 
Radial component 
\begin{equation}
B_\parallel(r)=B_\parallel(a)\left({a\over r}\right)^2
\end{equation}
due to the magnetic flux conservation, $B_\parallel d\sigma={\rm const}$. 

\begin{figure}
\centering
\includegraphics[width=8.8cm]{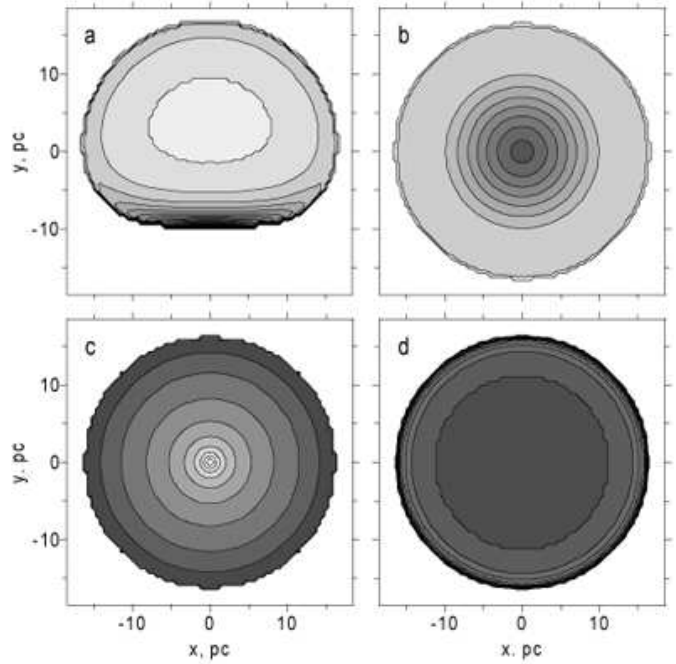}

\caption[]{{\bf a-d.} Logarithmic distributions of 
thermal X-ray surface brightness $S$ 
in ${\rm erg\ s^{-1}\ cm^{-2}\ st^{-1}}$ 
for photon energy $\varepsilon>0.1\ {\rm keV}$ ({\bf a, b}), 
and radio surface brightness $S_\nu$ at some frequency 
in relative units ({\bf c, d}). 
The SNR model is $b$, power $s=2$. Angles $\delta$ and $\phi$ are shown 
in the figure. 
{\bf a, b}:~$\lg S_{\max}=-3.1$, 
 $\Delta \lg S=0.3$. 
{\bf c}:~
$\Delta \lg S_\nu=0.3$. 
{\bf d}:~
$\Delta \lg S_\nu=0.15$. 
The darker colour represents a higher intensity. 
The arrow indicates a magnetic field orientation. 
            }
\label{fig-1}
\end{figure}

In each fluid element, energy density $\omega$ of relativistic 
particles is proportional to the energy density of the magnetic field 
\begin{equation}
\omega\equiv\int\limits^{E_{\max}}_{E_{\min}}EN(E)dE=
K\int\limits^{E_{\max}}_{E_{\min}}E^{1-s}dE
\propto B^2,
\end{equation}
which yields 
$K\propto B^2\big(E_{\max}^{2-s}-E_{\min}^{2-s}\big)^{-1}$ 
for $s\neq 2$ and 
$K\propto B^2\ln \big(E_{\min}/E_{\max}\big)$ for $s=2$. 
We are to expect the variation of $E_{\max}$ and $E_{\min}$ over the 
surface of a nonspherical SNR and downstream. 
However, the maximum energy, to which particles can be accelerated, 
only varies few times during the whole adiabatic phase 
(R98). The ISM non\-u\-ni\-for\-mi\-ty does not affect the evolution of 
SNR at the previous free expansion stage. Thus, we assume that 
possible variations of the minimum and maximum electron energies over the 
remnant's surface are likely to be minute 
and, in the first approach, we may neglect the surface 
variation of $E_{\max}$ and $E_{\min}$ caused by a nonuniform 
medium\footnote{Since $\omega_{\rm s}\propto P_{\rm s}$ (Reynolds \& 
Chevalier 
\cite{Rey-Chev-81}) and shock velocity $D\propto (\rho^o_{\rm s}R^3)^{-1/2}$ 
in a nonuniform medium (Hnatyk \cite{Hn88}), 
the variation of energy density 
$\omega_{\rm s}\propto R^{-3}$ over the surface 
lies within factors $3$ to $9$ for models $a$-$c$.}. 

Each individual electron loses its energy due to the adiabatic expansion: 
\begin{equation}
\dot{E}=E\ {\dot{\overline{\rho}}\over 3\overline{\rho}}, 
\label{en-loss}
\end{equation}
where $\overline{\rho}(r)=\rho(r)/\rho_{\rm s}$ (R98)  
and, therefore, $E(r)\propto\overline{\rho}(r)^{1/3}$ downstream. 
Thus, 
\begin{equation}
K(r)\propto B(r)^2\overline{\rho}(r)^{(s-2)/3}. 
\end{equation}

Radio morphology depends on aspect angle $\phi$ 
between the line of sight and the ambient magnetic field 
(R98), also on inclination angle $\delta$ between 
the density gradient and the plane of the sky and, 
in a complex 3-D case, on the third angle between the 
density gradient and the magnetic field. 

\subsection{Results}

Fig.~\ref{fig-1}a-b demonstrates the influence of the projection on a 
thermal X-ray morphology of SNR. The X-ray brightness maximum 
located near the shock 
front in the shell-like projection ($\delta=0\degr$) moves towards 
the centre 
of the projection with the increase of $\delta$ from $0\degr$ to $90\degr$. 
For clarity we only consider here the most emphatic limit case 
$\delta=90\degr$. 
Radio images (Fig.~\ref{fig-1}c-d) 
show that the radio limb-brightened morphology 
clearly appears at $\phi=0\degr$, i.~e. if both the density 
gradient and the magnetic field are nearly aligned. 

Variation of the magnetic field orientation changes the 
radio morphology from shell-like to barrel-like 
(Kes\-te\-ven \& Caswell \cite{Kest-Casw-87}, Gaensler \cite{Gaensler-98}). 
Contrast $\lg (S_{\nu,\max}/S_{\nu,\min})$ 
in the radio surface brightness decreases with increasing 
$\phi$, from  $2.7$ ($\phi=0\degr$) 
to 1.8 ($\phi=90\degr$). 
Such behaviour of the radio morphology 
may be used for testing orientation of the 
magnetic field. 

Thus, we found that the morphological properties of the projected composites 
match the basic features of 
the TXC class: centrally-peaked distribution of the thermal X-ray surface 
brightness is within the area of the radio shell; emission arises from the 
swept-up ISM material. 

\begin{figure}
\centering
\includegraphics[width=8.8cm]{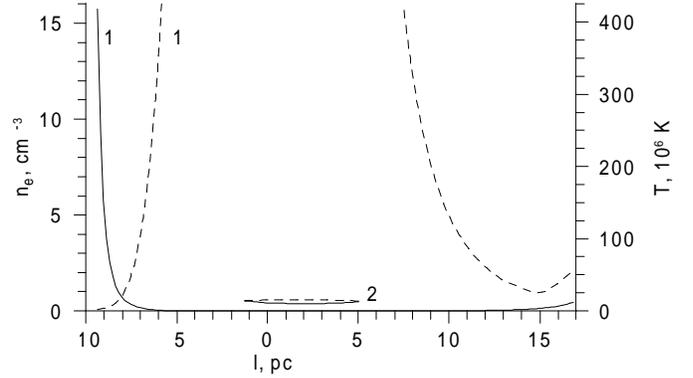}

\caption[]{Distribution of density (solid curves) and 
temperature (dashed curves) along the line of sight inside SNR 
(model {\it b}, $\delta=90\degr$): 1 -- at the center of 
the projection, 2 -- at $0.95R_{\rm p}$.  
Emission measure $\int n_e^2dl=74\ {\rm cm^{-6}pc}$ at the center 
and $\int n_e^2dl=1.1\ {\rm cm^{-6}pc}$ at $0.95R_{\rm p}$.
           }
\label{fig-add}
\end{figure}

Let us consider physical properties of TXCs. 
a) The column number density increases from the edge towards 
the centre of the 
projection (e.g., for model $b$ from $10^{18.9}\ {\rm cm^{-2}}$ to 
$10^{19.4}\ {\rm cm^{-2}}$). 
b) The diffuse optical nebulosity over the 
internal region of the projection may naturally take place in such a 
model. 
c) Emission measure $\int n_e^2dl$ 
($n_e$ is the electron number density, 
$l$ is the length within SNR) 
is the highest in the X-ray peak 
because both $n_e$ and $l$ are maximum there 
(Fig.~\ref{fig-add}). 

\begin{figure}[t]
\centering
\includegraphics[width=8.8cm]{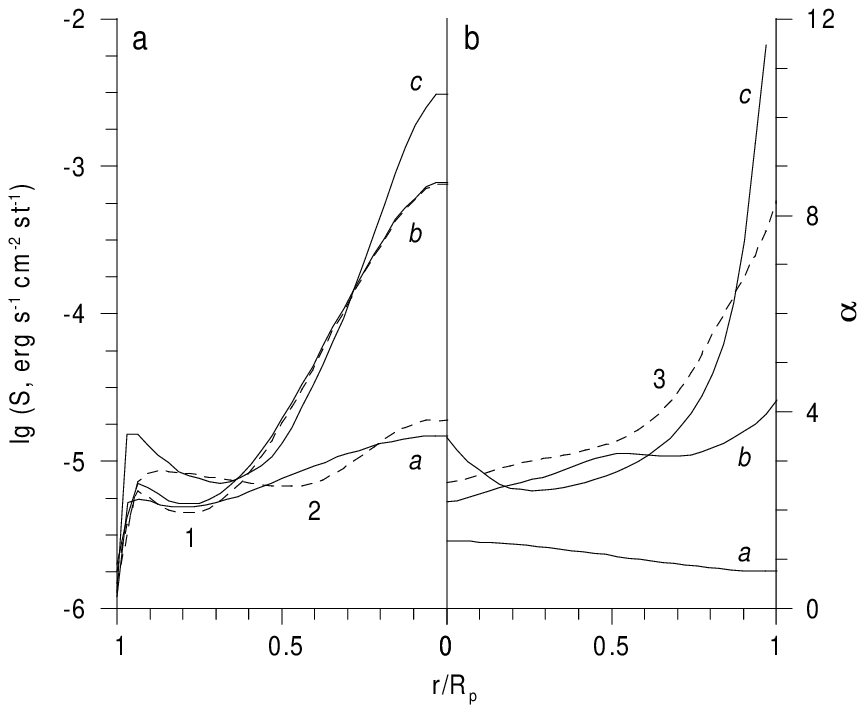}

\caption[]{{\bf a and b.} Evolution of the distribution of thermal X-ray 
surface brightness ({\bf a}) and spectral index ({\bf b}). 
Solid curves are labelled with the model codes according to 
Table~\ref{tabl-1}; they represent $\lg S$ in band 
$\varepsilon>0.1\ {\rm keV}$ and $\alpha$ at $5\ {\rm keV}$. 
Dashed lines represent model $b$ in other bands: 
1 -- $S$ in $\varepsilon=0.1-2.4\ {\rm keV}$, 
2 -- $S$ in $\varepsilon>4.5\ {\rm keV}$ (multiplied by $10^2$), 
3 -- $\alpha$ at $10\ {\rm keV}$. 
Radii are normalized to unity. 
           }
\label{fig-2}
\end{figure}

As Fig.~\ref{fig-2} demonstrates, the distribution of X-ray surface 
brightness has strong maximum $S_{\rm c}$ around the centre and 
a weaker shell with second maximum $S_{\rm max,\ \!2}$ just behind the 
forward shock. 
It is essential that such a morphology takes place in different X-ray bands 
(lines b, 1, 2). 
The contrasts $S_{\rm c}/S_{\rm max,\ \!2}$ in X-ray surface brightness 
depend on the photon energy band and may lie 
within a wide range: in our models from 3 to 200 
for $\varepsilon>0.1\ {\rm keV}$ (Table~\ref{tabl-1}). 
The ratios of X-ray luminosity 
$\int S(r)2\pi rdr$ of central region 
$R<0.2R_{\rm p}$ to the luminosity 
beyond $R>0.6R_{\rm p}$ are 0.16, 5.2 and 16 in models 
$a$, $b$ and $c$, respectively.  
Thus, observational property d of TXCs takes place 
just at the adiabatic stage. 

Surface distribution of spectral index 
$\alpha=\partial\ln P_{\rm c}/\partial\ln\varepsilon$, 
of the thermal X-ray emission 
where $P_{\rm c}$ is the continuum emissivity 
and $\varepsilon$ is the photon energy, 
gives us profiles of 
effective temperature $T$ of the column of emitting gas 
($\alpha\propto T^{-1}$, if the Gaunt factor is assumed to be constant). 
Fig.~\ref{fig-2} 
shows that the temperature may either increase or decrease towards the 
centre. Decreasing takes place early in the adiabatic phase.  
Variation of the spectral index lies within factors $1.6$ to $3.6$ 
at the adiabatic stage (Table~\ref{tabl-1}); the contrast 
in the spectral index distribution increases with age. Such  
values correspond to the possible range of temperature variation 
over the projection of thermal X-ray composites. 

\begin{figure}
\centering
\includegraphics[width=8.8cm]{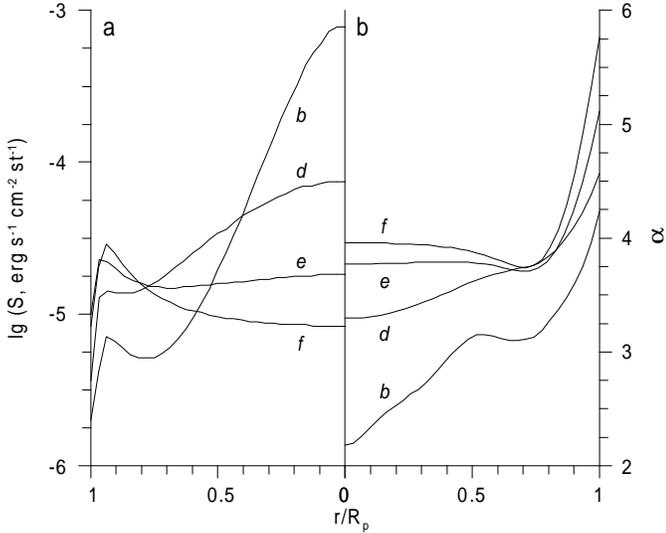}

\caption[]{{\bf a and b.} Influence of $h$ on the 
distribution of thermal X-ray 
surface brightness ($\lg S^{>0.1\ {\rm keV}}$) ({\bf a}) and 
spectral index ($\alpha^{5\ {\rm keV}}$) ({\bf b}). 
Curves are labelled with the model codes according to Table~\ref{tabl-1}. 
Radii are normalized to unity. 
           }
\label{fig-3}
\end{figure}

In order to reveal the dependence of the distributions of $S$ and $\alpha$ 
on the ISM density gradient, 
a number of models with different $h$ were calculated (Fig.~\ref{fig-3} 
and Table~\ref{tabl-1}). 
The surface brightness distribution has a stronger peak for a stronger gradient. 
With increasing $h$, the outer shell becomes more prominent in the projection. 
Only a scale-height of order $h<10\ {\rm pc}$ can cause projected 
composites. A less strong gradient of the ambient density makes 
effective temperature $T$ 
more uniformly distributed in the internal part of the projection. 

\section{Discussion}

\subsection{Projection model versus other models} 

Distinctions of the previous physical models for TXC 
from the projection model are noted below. 
1) The model with strong cooling can be used 
within the frame of the radiative model of SNR; the 
presented model describes TXCs as SNRs in the adiabatic phase of 
their evolution. 
2) In comparison with the projection model, the spectrum of the central 
region in the model with thermal conduction is softer due to reducing 
the temperature (Jones et al. \cite{Workshop-97}). 
3) The models with thermal conduction or evaporation increase the density in 
the internal part of SNR. The projection model 
does not require such modification of internal density distribution; 
the density profiles in this model are similar to those in the Sedov 
(\cite{Sedov}) solution. 
4) A small-scale inhomogeneous ISM is required for the model with evaporation. 
Projected composites are a consequence of large-scale nonuniformity of 
the ISM with the scale-height of order $<10\ {\rm pc}$. 
5) Other possibilities of creating a centrally-filled morphology, such as  
differential absorption 
or emission from ejecta, specifically modify 
the spectra of the object. 

\subsection{Magnetic field and density gradient 
		orientations in molecular clouds}

The projection model of TXCs assumes that an SNR evol\-ves in 
a nonuniform medium with a different nature of nonuniformity 
and does not require the presence of a molecular cloud near SNR. 
However, since TXCs are mainly located near such clouds, 
we have considered here evolution of SNR at the edge of a cloud, as 
the most probable case. 
Magnetic field and ISM density gradient orientations 
are critical parameters for the proposed model. Therefore, let us 
find observations which reveal a  
close alignment between the 
density gradient and the magnetic field in molecular clouds. 
This might support 
the presented model since SNR is most clearly TXC under such an 
alignment. 

The observational techniques used to measure the orientation and strength 
of the magnetic field in molecular clouds are reviewed by 
Troland (\cite{Troland-90}) and Crutcher (\cite{Crutcher-1994}). 
Zeeman effect observations give information about the strength of 
magnetic field component $B_{\rm los}$ parallel to the line of sight 
(Crutcher et al. \cite{Crutcher-et-al-93}). 
To define the orientation of magnetic field component $B_{\rm pos}$  
in the plane of the sky, polarization measurements of 
the radio, 
infrared and optical line emission 
are used. Polarization data 
maps reveal the clouds' magnetic field morphology 
in the plane of the sky 
but do not enable us to find the amplitude of $B_{\rm pos}$. 
Thus, we cannot obtain the actual three-dimensional orientation of 
the magnetic field from the observations because only one component of 
$B$ can be measured directly. 

It is also difficult to find out the density gradient direction 
in a cloud, especially in low density regions 
($\sim 10\div100\ {\rm cm^{-3}}$) which we are interested in. 
Column density maps yield a projected 2-D structure of a 
3-D cloud. Since the density distribution along the line of sight in a cloud 
generally remains unknown, we cannot draw conclusions about a 
three-dimensional direction of the density gradient. Another complication 
is a mostly filamentary stucture of clouds which prevents us from seeing 
a large-scale gradient of density. 

For our purpose, it is reasonable to look for 
clouds with $B_{\rm los}\approx 0$ when 
polarization directions show a morphology of total field $B$. 
There are 27 clouds with measurements of the Zeeman effect (Crutcher 
\cite{Crutcher-99}). 
Crutcher et al. (\cite{Crutcher-et-al-93}) have reported 
10 sensitive observations without any detections 
of the Zeeman splitting. Since Zeeman observations are only sensitive to 
$B_{\rm los},$ the authors assume that the magnetic fields in these clouds 
lie mostly in the plane of the sky%
. 
The cloud positions support this assumption, for 
looking at 9 of these clouds we are looking nearly 
perpendicularly at the local spiral arm, where the field is predominantly 
directed along the arm (Crutcher et al. \cite{Crutcher-et-al-93}). 

One of the most interesting cases of these observations is the 
nearby ($\sim140\ {\rm pc}$) Taurus complex 
for which several polarization analyses have been made. 
There is additional support for $B_{\rm los}\approx 0$ in it 
(at least around the TMC-1C core): this  
is a high aspect ratio of the core flattening along 
$B_{\rm pos}$ (Troland et al. \cite{Troland-et-al-1996}). 
Thus, we may assume that the map of polarizations in the Taurus cloud 
(e.~g. Moneti et al. \cite{Moneti-et-al-84}) 
gives us a large-scale direction of $B$. 

Denser regions in Taurus are filamentary in the plane of the sky, as the 
column density 
map shows (e.~g. Wiseman \& Adams \cite{Wiseman-Adams-94}), but 
we expect that a large-scale ($\sim 10\ {\rm pc}$) gradient of density 
in the Taurus cloud should be nearly aligned with the direction of 
magnetic field. The cloud has a flattened structure which is belived to be 
the result of the collapse controlled by interstellar magnetic field 
(Heyer et al. \cite{Heyer-et-al-87}). 
In such a case, gas tends to collapse along magnetic field lines 
and this causes the minor axis of the cloud to be 
parallel to $B\approx B_{\rm pos}$. 
The direction of the density gradient should be along the minor axis 
of the collapsed object. 

We would like to note that in a situation when 
we cannot have firm orientations of magnetic field and density 
gradient from observations, SNR itself may sometimes be 
considered as a test of these orientations 
because the radio morphology of SNR depends on the magnetic field 
and the thermal X-ray morphology -- on the interstellar density. 

\subsection{Barrel-like TXCs}

It is obvious that the projection model of TXCs does not 
require the density gradient to be strictly along the line of sight. 
When inclination angle $\delta$ changes from $90\degr$ to $0\degr$, the 
thermal X-ray peak moves from the center of the projection to an edge. 
Therefore, the boundary between SNRs which are either centrally-peaked 
or limb-brightened in X-rays is not quite clear 
from the point of view of the proposed model: the both cases are the same 
shell-like SNRs projected onto the plane of the sky in different ways. 

Another interesting fact concerns a more crucial component of the 
model. The magnetic field component along the line of sight should be 
maximum among other components to provide a shell-like radio morphology. 
As it has been noted above, if the magnetic field is oriented primarily 
in the projection plane, we can observe a barrel-like morphology 
(Fig~\ref{fig-1}). 
Such a situation suggests looking for remnants which 
have a barrel-like radio morphology ($\phi$ close to $90\degr$) coupled with 
a centrally-brightened thermal X-ray one. Such remnants yield 
additional support for the projection model of TXCs, since these SNRs 
are simply another case of the magnetic field orientation. 

The list of 17 SNRs which are bilateral in the radio band is presented by 
Gaensler (\cite{Gaensler-98}). Unfortunately, X-ray observations 
are only known for 6 of them (Green \cite{Green-2000}). 
Three of these 6 SNRs are of a shell-like type, both in radio and X-rays
(\object{$\gamma$~Cygni}, \object{G156.2+5.7} and \object{SN 1006}), 
two others (\object{G296.5+10.0} and  \object{RCW89}) have pulsars. 
Only \object{VRO~42.05.01} is centrally-brightened in thermal X-rays 
(Burrows \& Guo \cite{Burrows-Guo-94}; Guo \& Burrows \cite{Guo-Burrows-97}). 
It, therefore, is a candidate for a "barrel-like TXC". 
The results of future X-ray observations will show whether there exist 
more such remnants. 

\subsection{$\gamma$-rays from TXCs}

Gamma-ray emission from SNRs is important because 
it allows us to draw conclusions about the cosmic 
ray acceleration on shocks.  

Recent observations of nonthermal X-rays from the 
\object{SN~Tycho} (Ammosov et al. \cite{Ammosov-et-al-94}), 
\object{SN~1006} (Koyama et al. \cite{Koyama_et_al-95}), 
\object{Cas~A}  (Allen et al. \cite{Allen_et_al-97}), 
\object{G347.3-0.5} (Koyama t al. \cite{Koyama_et_al-97}), 
\object{IC~443} (Keohane et al. \cite{Keohane-P-97}), 
\object{G266.2-1.2} (Slane et al. \cite{Slane-Huges-et-al-00}) 
and TeV $\gamma$-rays 
from \object{SN~1006} (Tanimori et al. \cite{Tanimori_et_al-98}), 
and \object{G347.3-0.5} (Muraishi et al. \cite{Muraishi-et-al-00}) 
give firm experimental confirmations that Galaxy 
cosmic rays are accelerated on the shocks of SNRs up to the 
energies $10^{14}\ {\rm eV}$. 

The third EGRET catalog (Hartman et al. \cite{3EG}) 
lists 170 unidentified GeV $\gamma$-ray sources. 
74 of them are located at $|b|<10\degr$ 
and 22 of these sources coinside with directions 
towards known SNRs 
(Romero et al. \cite{Romero-B-T-99}). 6 SNRs from this list are TXCs
(\object{IC~443}, \object{MSH~11-61A}, \object{W28}, \object{W44}, 
\object{3C396}, \object{G359.1-0.5}),  
2 are Crab-like (\object{CTB~87}, 
\object{G27.8+0.6}) and 4 are of a shell-like type (\object{Puppis~A}, 
\object{Vela}, \object{G359.0-0.9}, 
\object{$\gamma$-Cygni}). It is not clear which morphological class 
other SNRs belong to, because no X-ray observations of them have been 
reported (Green \cite{Green-2000}).  

Different emission mechanisms compete in the analysis of 
observed $\gamma$-ray spectra. 
Unfortunately, we still have no direct observational confirmations as to 
proton acceleration in SNRs. Only $\gamma$-rays from $\pi^o$ meson 
decays created in 
proton-nucleon interactions allow us to look inside the 
cosmic ray nuclear component 
acceleration processes. 
To make proton origin $\gamma$-rays dominating in an SNR model, 
we need a high number density of target nuclei 
($\sim 10^2-10^5\ {\rm cm^{-3}}$). 
Therefore, $\pi^o$ decay $\gamma$-rays are expected  
from SNRs which interact with molecular clouds.
The new model for TXCs 
strongly suggests: the thermal X-ray peak inside the radio shell of SNR 
testifies that one part of the SNR shock enters a denser medium 
than other parts of the shell. Thus, the proposed model of TXCs allows  
us to consider members of this class as prospective sources of $\pi^o$ 
decay $\gamma$-emission. 

\section {Conclusions}

Considering 2-D or 3-D models of SNRs it is necessary to take into 
account the effects of the projection. 
Once projected onto the plane of the sky, 
such an object changes its appearance 
depending on the actual density contrast across the remnant and on 
the angle 
between the density gradient and the direction towards the observer. 
If the ambient density gradient does not lie in the plane of the projection 
and is strong enough, and 
if the magnetic field is nearly aligned with the line of sight, 
then the 
visible thermal X-ray morphology of SNR will be centrally-filled, while the 
radio morphology will remain limb-brightened. 
The projection effect is maximum when the density gradient is oriented 
along the line of sight. 
Only a scale-height of order $h<10\ {\rm pc}$ 
in ambient medium density distribution 
can cause projected composites. 

All theoretical properties of projected composites 
correspond to 
observational properties of thermal X-ray composites. 
Thus, the circumstances should exist when SNRs are projected as 
centrally-filled X-ray objects. 

Majority of the members of a mixed-morphology class are really located 
near molecular clouds. 
Therefore, at least a part of them may be the result 
of a simple projection effect of the adiabatic SNR evolved in a 
nonuniform medium, e.g., at the edge of a molecular cloud.

If the prediction that a part of barrel-like radio SNRs have a 
centrally-filled thermal X-ray morphology is confirmed, it will provide 
additional support for the proposed 
model. The model suggests that TXCs may be prospective sources of 
proton origin $\gamma$-rays. 





\begin{thebibliography}{}

\bibitem[1997]{Allen_et_al-97}
 Allen G. E., Keohane J. W., Gotthelf E. V., et al., 1997, ApJ 487, L97
\bibitem[1994]{Ammosov-et-al-94}
 Ammosov et al., 1994, Sov. Astron. Lett. 20, 191
\bibitem[1998]{Blondin-et-al-98}
 Blondin et al., 1998, ApJ 500, 342
\bibitem[1994]{Burrows-Guo-94}
 Burrows D. \& Guo Z., 1994, ApJL 421, L19
\bibitem[1999]{W44}
 Cox D., Shelton R., Maciejewski W., et al., 1999, ApJ 524, 179 
\bibitem[1994]{Crutcher-1994}
 Crutcher R., 1994, in ASP Conf. Ser. 65, Clouds, Cores, and Low Mass Stars, 
 ed. D. Clemens \& R. Barvainis (Bo\-ok\-Craf\-ters, Inc.), 87
\bibitem[1999]{Crutcher-99}
 Crutcher R., 1999, ApJ 520, 706
\bibitem[1993]{Crutcher-et-al-93}
 Crutcher R., Troland T., Goodman A. et al., 1993, ApJ 407, 175
\bibitem[1998]{Gaensler-98}
 Gaensler B., 1998, ApJ 493, 781 
\bibitem[1990]{Goodman-et-al-90}
 Goodman et al., 1990, ApJ 359, 363
\bibitem[2000]{Green-2000}
 Green D.A., 2000, `A Catalogue of Galactic Su\-per\-no\-va Rem\-nants (2000 August
        ver\-sion)', Mul\-lard Ra\-dio Astronomy Observatory, Cavendish La\-bo\-ra\-to\-ry,
        Cam\-b\-rid\-ge, United Kingdom 
	(
	\verb"http://www.mrao.cam.ac.uk/surveys/snrs/")
\bibitem[1997]{Guo-Burrows-97}
 Guo Z. \& Burrows D., 1997, ApJ 480, L51 
\bibitem[1998]{Workshop-97}
 Jones T., Rudnick, L., Jun, B.-I., et al., 1998, PASP 110, 125 
\bibitem[1999]{3EG}
 Hartman, R. C., et al., 1999, ApJS, 123, 79
\bibitem[1987]{Heyer-et-al-87}
 Heyer M., Vrba F., Snell R. et al., 1987, ApJ 321, 855
\bibitem[1988]{Hn88}
 Hnatyk B., 1988, Sov. Astron. Lett. 14, 309 
\bibitem[1998]{Hn-Pet-98}
 Hnatyk B., \& Petruk O., 1998, Condensed Matter Physics 1, 655 
	[available also as preprint astro-ph/9902158].
\bibitem[1999]{Hn-Pet-99}
 Hnatyk B., Petruk O., 1999, A\&A 344, 295 
\bibitem[1997]{Keohane-P-97}
 Keohane J. W., Petre R., Gotthelf E., et al., 1997, ApJ 484, 350
\bibitem[1987]{Kest-Casw-87}
 Kesteven M., Caswell J., 1987, A\&A 183, 118 
\bibitem[1995]{Koyama_et_al-95}
 Koyama K., Petre R., Gotthelf E. V, et al., 1995, Nature 378, 255
\bibitem[1997]{Koyama_et_al-97}
 Koyama K., Kinugasa K., Matsuzaki K., et al., 1997, PASJ 49, L7
\bibitem[1991]{Long-et-al-91}
 Long K., Blair W., White R., Matsui Y., 1991, ApJ 373, 567
\bibitem[2000]{Matthews-Wilson-2000}
 Matthews B. \& Wilson C., 2000, ApJ 531, 868
\bibitem[1997]{Messinger-et-al-97}
 Messinger D., Whittet D., Roberge W., 1997, ApJ 487, 314
\bibitem[1984]{Moneti-et-al-84}
 Moneti A., Pipher L., Helfer H. et al., 1984, ApJ 282, 509
\bibitem[2000]{Muraishi-et-al-00}
 Muraishi H., Tanimori T., Yanagita S., et al., 2000, A\&A 354, L57
\bibitem[1977]{Raym-Smith77}
 Raymond J., Smith B., 1977, ApJS 35, 419
\bibitem[1998]{Reyn-98}
 Reynolds S., 1998, ApJ 493, 375 (R98)
\bibitem[1981]{Rey-Chev-81}
 Reynolds S., Chevalier R., 1981, ApJ 245, 912
\bibitem[1996]{Rho-Petre-96}
 Rho J., Petre R., 1996, ApJ 467, 698 
\bibitem[1998]{Rho-Petre-98}
 Rho J., Petre R., 1998, ApJ 503, L167 (RP98)
\bibitem[1994]{Rho-et-al-94}
 Rho J., Petre R., Schlegel E., 1994, ApJ 430, 757
\bibitem[1999]{Romero-B-T-99}
 Romero G., Benaglia P., Torres D., 1999, A\&A 348, 868
\bibitem[1959]{Sedov}
 Sedov L., 1959, Similarity and Dimensional Methods in Me\-cha\-nics. 
  Academic, New York
\bibitem[1990]{Seward-catalog}
 Seward F., 1990, ApJS 73, 781
\bibitem[1999]{W44-II}
 Shelton R., Cox D., Maciejewski W., et al., 1999, ApJ 524, 192 
\bibitem[2000]{Slane-Huges-et-al-00}
 Slane P., Hughes J., Edgar R., et al., 2000, astro-ph/0010510
\bibitem[1999]{SunWangChen-99} 
 Sun M., Wang Z., Chen Y., 1999, ApJ 511, 274
\bibitem[1998]{Tanimori_et_al-98}
 Tanimori T., Hayami Y., Kamei S., et al., 1998, ApJ 497, L25
\bibitem[1990]{Troland-90}
 Troland T., 1990, in IAU Symp. 140,Galactic and Intergalactic Magnetic 
 Fields, ed. R. Beck, P. Kronberg \& R. Wielebinski (Dordrecht: Kluwer), 293 
\bibitem[1996]{Troland-et-al-1996}
 Troland T., Crutcher R., Goodman A. et al., 1996, ApJ 471, 302
\bibitem[1996]{MOST-catalog}
 Whiteoak J.B., Green A., 1996, A\&AS 118, 329 
\bibitem[1994]{Wiseman-Adams-94}
 Wiseman J. \& Adams F., 1994, ApJ 435, 708

\end{thebibliography}
\end{document}